\documentclass[conference]{IEEEtran}
\IEEEoverridecommandlockouts
\usepackage{lipsum,booktabs}
\usepackage{comment}
\usepackage{amsmath,amssymb,amsfonts}
\usepackage{algorithmic}
\usepackage{mathtools, cuted}
\usepackage{graphicx}
\usepackage{hyperref}
\usepackage{orcidlink}
\usepackage{textcomp}
\usepackage{xcolor}
\usepackage{balance}
\usepackage{enumitem}
\usepackage{pifont}
\def\passed{\ding{52}}
\def\failed{\ding{55}}

\usepackage{doi}

\usepackage{multibib}

\newcites{article}{Article References}
\newcites{software}{Software References}

\newcommand{\turntitle}[1]{\rotatebox{90}{ \textbf{#1} }}

\newcommand{\framedtitle}[1]{
  \parbox[c][4em][c]{0.065\hsize}{
   {\centering #1}
  }
}

\newcommand{\softname}[1]{\textsc{#1}}
\newcommand{\readme}{README}

\begin{document}

\title{FAIRSECO: An Extensible Framework for Impact Measurement of Research Software}

\author{\IEEEauthorblockN{Deekshitha\orcidlink{0000-0003-1831-8941}\IEEEauthorrefmark{1}\IEEEauthorrefmark{2}\IEEEauthorrefmark{3},
Siamak Farshidi\orcidlink{0000-0003-3270-4398}\IEEEauthorrefmark{1},
Jason Maassen\orcidlink{0000-0002-8172-4865}\IEEEauthorrefmark{2}, Rena Bakhshi\orcidlink{0000-0002-2932-3028}\IEEEauthorrefmark{2},
Rob van Nieuwpoort\orcidlink{0000-0002-2947-9444}\IEEEauthorrefmark{2}\IEEEauthorrefmark{3},
Slinger Jansen\orcidlink{0000-0003-3752-2868}\IEEEauthorrefmark{1}
}
\IEEEauthorblockA{\IEEEauthorrefmark{1}Department of Information and Computing Sciences, Utrecht University, Utrecht, the Netherlands
    \\\{d.deekshitha,s.farshidi,slinger.jansen\}@uu.nl}
\IEEEauthorblockA{\IEEEauthorrefmark{2}Netherlands eScience Center, Amsterdam, the Netherlands
    \\\{j.maassen,r.bakhshi,r.vannieuwpoort\}@esciencecenter.nl}
\IEEEauthorblockA{\IEEEauthorrefmark{3}University of Amsterdam, Amsterdam, the Netherlands
    }}
    
\maketitle 

\begin{abstract}
 % Reusing research software, the software that advances society by supporting and contributing to academic research, is hard. The difficulty lies in locating the necessary algorithm, importing the relevant methods, and giving proper recognition to the original developers of the research software. To that end, we present FAIRSECO, an extensible framework %to make research software more FAIR, i.e., Findable, Accessible, Interoperable, and Reusable. The FAIRSECO framework infrastructure  that addresses two critical information needs: firstly, it provides potential users of research software with metrics related to software quality, provenance information, and impact data. Secondly, the infrastructure provides information for those who wish to measure the success of a project by offering impact and citation data. If research software engineers start using frameworks such as FAIRSECO, they better measure and enlarge the impact of their research software.

 The growing usage of research software in the research community has highlighted the need to recognize and acknowledge the contributions made not only by researchers but also by Research Software Engineers. However, the existing methods for crediting research software and Research Software Engineers have proven to be insufficient. In response, we have developed FAIRSECO, an extensible open source framework with the objective of assessing the impact of research software in research through the evaluation of various factors. The FAIRSECO framework addresses two critical information needs: firstly, it provides potential users of research software with metrics related to software quality and FAIRness. Secondly, the framework provides information for those who wish to measure the success of a project by offering impact data. By exploring the quality and impact of research software, our aim is to ensure that Research Software Engineers receive the recognition they deserve for their valuable contributions.
 
\end{abstract}

\begin{IEEEkeywords}
FAIR, research software engineering, software impact measurement, software citations
\end{IEEEkeywords}

%%%%%%%%%%%%%%%%%%%%%%%%%%%%%%%%%%%%%%%%%%%%%%%%%%%%%%%%%%%%%%%%
\section{Introduction} \label{introduction}
The FAIR4RS Working Group~\citearticle{fair4wg} defines research software (RS) as source code files, algorithms, scripts, computational workflows and executables created during the research process or for a research purpose. In contrast, other types of software, such as operating systems, libraries, dependencies, and packages are not explicitly created for research but are employed in research activities. 

Recent studies~\citearticle{barker_michelle_2021_5762703,carver-2022} have shown that 33\% of international research produces new code and 90-95\%  of UK and US researchers use and acknowledge RS as important for their research. In many research projects, Research Software Engineers (RSEs) closely collaborate with researchers to understand their challenges and to develop RS that helps to provide the answers to their research questions~\citearticle{rse}. 
Reusing (parts of) RS is beneficial to the research community. It saves time and effort and encourages collaboration and good coding practices. By encouraging RS reuse, the developers increase their work's impact and foster a collaborative environment that drives scientific progress~\citearticle{istrate2022large}. 

Two factors contribute to the success of RS reuse. First, like research data, the RS should be made FAIR~\citearticle{barker2022introducing,katz2021taking}. Following the FAIR principles increases the potential for the RS to be found and reused by other researchers, as well as encouraging good coding practices and enabling software citation. 

Second, when reusing RS, proper credit should be given to the developers. Similar to scholarly papers and data sets, RS should get recognition through citations. The current lack of consistent software citation makes it difficult to measure the reuse and impact of RS and creates challenges in giving credit to RSEs for their contributions~\citearticle{van2016lightning}. Properly citing RS will assist the authors in obtaining financial support for further software development, thereby improving the RSEs career paths.

Many guidelines and tools exist that help developers improve the FAIRness, quality, and citeability of their code~\citearticle{Gomez_Diaz_2019,alliez_attributing_2020},~\citesoftware{Spaaks_howfairis_2022,Druskat_Citation_File_Format_2021,Verhoeven_tortellini_2021}. Currently, however, there is no single framework RSEs can use to get an overview of the FAIRness, quality and impact of their RS. Moreover, Gomez-Diaz and Recio~\citearticle{Gomez_Diaz_2019} and Istrate et al.~\citearticle{impact:22} argue that there is currently no sufficient method to evaluate the quality and impact of RS.
To address this issue, we have created the FAIR Research Software Ecosystem (FAIRSECO) framework~\citesoftware{Aydin_FAIRSECO_2023}. 
The FAIRSECO framework is designed to combine different RS metrics under one extensible framework. 
It combines data from many existing tools that provide information on license conflicts, dependencies, method-level code-reuse, bibliometric analysis, citation metadata, FAIRness aspects, code indexing, and Software Bill of Materials (SBOM) generation to generate a concise overview of the quality, FAIRness, and impact of RS. FAIRSECO then combines the output of these existing tools into two scores, one for quality and one for impact. 
This enables RSEs to quickly gain insight into the recognition their software already receives and provide suggestions on how they can improve the quality and potential impact of their software.

The idea of the FAIRSECO framework has been introduced in a conceptual short paper~\citearticle{FAIRSECO-2022}. However, that short paper focuses only on impact measurement through method reuse. While method reuse is important, we have evolved our FAIR software ideas and significantly broadened our scope.

The contribution of current work is twofold:

    \begin{itemize}
      \item \textbf{Integration of quality, FAIRness and impact measurement tools into a single extensible framework}: 
    The FAIRSECO framework consolidates 
    a comprehensive collection of tools designed to measure various aspects of RS quality, FAIRness, and impact into a single, unified framework. This provides researchers with a convenient and efficient solution for measuring and assessing 
    these aspects of their RS.
      \item \textbf{Quantifying the quality and impact of RS based on key features}: We provide a novel Quality Score ($S_{quality}$) for assessing the quality of RS based on key features, incorporating FAIRness, license violations, maintainability, and documentation as factors in the equation. By considering these aspects together, we provide a robust and objective method for systematically measuring and quantifying RS quality. 
    Similarly, an Impact Score ($S_{impact}$) is introduced based on three factors: the number of citations, the number of reused methods, and the Quality Score.
    \end{itemize}
    
We organize this paper as follows. 
First, we show what other tools are available and how they compare to the FAIRSECO framework in the related work Section~\ref{related}. This includes a brief description of each tool used by the FAIRSECO framework. Next, a full description is provided of the FAIRSECO framework in Section~\ref{fairseco}. The FAIRSECO framework is then demonstrated in action in Section~\ref{toolaction} by its application to an existing RS project. The evaluation of five RS projects using FAIRSECO is presented in Section~\ref{evaluation}. Finally, the paper concludes with a summary and discussion of the FAIRSECO framework infrastructure and a set of steps to take in the near future in Section~\ref{conclusion}.

\section{Related work}\label{related}
%position FAIRSECO among other tools (comparison)
%\section{Contributions}

We observed a shortage of tools and frameworks RSEs could use to measure their impact on the RS ecosystem. Therefore, we aimed to build an accurate framework for evaluating software quality and measuring the effect of RS.  To this end, we reviewed FAIRSECO framework and existing tools focusing on identification of license violation (license checking), examining the dependencies of a software project for outdated components or vulnerabilities (dependency checking), 
method level checking of code reuse (method-level checking), analysis of citation, impact related features (bibliometric analysis), 
checking presence of citation (citation file checking), checking compliance to the FAIR principles (FAIRness aspects), building the source code index (code indexing), and building a list of all the components and dependencies used in a software project (SBOM generation). 

\subsection{Tools of FAIRSECO}
\begin{table*}[!hbtp]
\scriptsize
\centering
\caption{Comparison Table: This table highlights a comparative analysis between FAIRSECO and other tools, emphasizing specific impact metrics.}
\label{tab:comparisonTable}
\begin{tabular}{|l|c|c|c|c|c|c|c|c|} 
\hline
\multicolumn{1}{|c|}{Tools} & \multicolumn{8}{c|}{Feature}                                   \\ 
\cline{2-9}
\multicolumn{1}{|c|}{}      & \framedtitle{License checking} & \framedtitle{Dependency checking} & \framedtitle{Method level checking} & \framedtitle{Bibliometric Analysis} & \framedtitle{Citation file checking} & \framedtitle{FAIRness Aspect} & \framedtitle{Code indexing} & \framedtitle{SBOM Generation}  \\ 
\hline
\softname{howfairis} & \failed & \failed & \failed  & \failed & \passed & \passed & \failed & \failed  \\ 
\hline
\softname{SearchSECO} & \failed & \failed & \passed & \failed & \failed & \failed & \passed &  \failed \\ 
\hline
\softname{Tortellini} & \passed & \failed   & \failed  & \failed & \failed & \failed & \failed & \failed  \\ 
\hline
\softname{Libraries.io} & \passed  & \passed& \failed   & \failed  & \failed  & \failed & \failed   & \failed     \\ 
\hline
\softname{swg-graph}   & \failed & \failed & \failed  & \failed & \failed   & \failed & \passed  & \failed \\ 
\hline
RSD & \failed  & \failed     & \failed & \passed  & \passed   & \passed   & \failed   & \failed    \\ 
\hline
\softname{Depsy} & \failed  & \failed     & \failed & \passed  & \failed   & \failed   & \failed   & \failed    \\ 
\hline
\softname{GitHub} & \failed  & \passed     & \failed & \passed  & \passed   & \failed   & \passed   & \failed    \\
\hline
\softname{GrimoireLab} & \failed  & \failed     & \failed & \passed  & \failed  & \failed   & \passed   & \failed    \\ 
\hline
\softname{SIRGRID} & \failed  & \passed     & \failed & \failed  & \failed   & \failed   & \failed   & \failed    \\ 
\hline
\softname{SQAaaS} & \passed  & \failed     & \failed & \passed  & \passed   & \passed   & \failed   & \failed    \\ 
\hline
FAIRSECO    & \passed   & \failed    & \passed  & \passed    & \passed   & \passed   & \passed   & \passed  \\ 
\hline
\end{tabular}
\end{table*}

We integrated several tools and external data sources into the FAIRSECO framework, namely, the tools \softname{howfairis}, \softname{Tortellini}, and \softname{SearchSECO}. 

%A brief description of each tool used in comparison with the FAIRSECO framework is provided in the following paragraphs.

\softname{howfairis} is a Python package to analyze a GitHub or GitLab repository's compliance with the recommendations given on fair-software.eu ~\citesoftware{Spaaks_howfairis_2022}. The \softname{howfairis} generates a FAIRness report with the help of \softname{fairtally}~\citesoftware{Verhoeven_fairtally_2021}. Based on this report, users identify areas where improvements are needed to make their repository more compliant.

The Netherlands eScience Center has developed a tool called \softname{Tortellini}~\citesoftware{Verhoeven_tortellini_2021}, a GitHub action that checks for any licensing issues in a given software, such as incompatible licenses in the software's dependencies.

\softname{SearchSECO}~\citearticle{jansen2020searchseco} is a hash-based index for code fragments that enables searching for source code at the method level in the global software ecosystem~\citearticle{jansen2015scientists}. \softname{SearchSECO} supports a number of languages and deals with multi-language projects. 
%SearchSECO is a language-agnostic search engine and research platform developed by the Utrecht University SecureSECO team.
%It enables searching source code at the method level in the worldwide software ecosystem~\citeofsoftware{Jansen_SearchSECO_-_A}. It provides finer-grained and more efficient searches than other search mechanisms, covers more of the
%software ecosystem, and offers mechanisms for source code provenance.
% ...

%SearchSECO is a language-agnostic search engine and research platform developed by the Utrecht University SecureSECO team. It enables searching source code at the method level in the worldwide software ecosystem~\citesoftware{Jansen_SearchSECO_-_A}. SearchSECO provides finer-grained and more efficient searches than other search mechanisms, and covers more of the software ecosystem, and offers mechanisms for source code provenance.

Table~\ref{tab:comparisonTable} summarizes the functionality of each of the tools described above. 
\softname{Howfairis} focuses on citation file checking and FAIRness aspect but lacks support for dependency checking and other tasks. Similarly, \softname{SearchSECO} only performs method-level checking and code indexing, and \softname{Tortellini} only supports license checking. Finally, FAIRSECO framework performs all listed tasks, except dependency checking. In particular, FARSECO performs license checking, method-level checking, citation file checking, FAIRness aspect checking, code indexing with the help of \softname{Howfairis}, \softname{Tortellini}, and \softname{SearchSECO}. 
%
%Finally, FAIRSECO combines the output of these tools to perform all listed tasks, including license checking, method-level checking, bibliometric analysis, citation file checking, FAIRness aspect checking, code indexing, and SBOM generation, except dependency checking
However, FAIRSECO is limited to the Github repositories.

\subsection{Related platforms}

\href{https://github.com}{GitHub}, the largest open-source repository, uses metrics such as stars, watchers, and forks (as a part of its Bibliometric Analysis) to measure the impact of software. The star system in GitHub functions similarly to the 'like' button in social media platforms, indicating user interest or support for a project or being used as a bookmark. However, there is a lack of comprehensive and well-founded empirical research to determine the exact meaning and practical implications of 'starring a project' in GitHub~\citearticle{borges2018s}. The credibility of the star metric on GitHub is questionable, as it is possible to create counterfeit or bot accounts that are used to give an excessive number of stars to a project. However, it performs dependency checking through dependency bots, and if citation file is present, it provides options for citing repository.

\softname{Libraries.io}~\citesoftware{libraries} is a platform designed to assist RSEs in discovering new open-source libraries, modules, and frameworks and keep track of the ones they rely on. Its primary objective is to enhance software quality by addressing three critical issues: discovery, maintainability, and sustainability. \softname{Libraries.io} performs dependency and license checking but does not support the other tasks. 

Depsy~\citearticle{depsy} measures impact of RS through software citations such as imports by other software and through references to software from papers. It is, however, limited to the RS available via the Python packages repository \href{https://pypi.org/}{PyPI} and the R packages repository \href{https://www.r-pkg.org/}{CRAN}.

The tooling and services known as Software Heritage Graph (\softname{swh-graph})~\citearticle{boldi2020ultra} offer quick access to the graph representation of the Software Heritage Archive, an archive that attempts to collect as much software as possible. Currently, it can claim to have the largest collection of source code. These services are collectively called \softname{swh-graph} and operate based on an in-memory, compressed representation of the Software Heritage Merkle DAG. \softname{swh-graph} is limited to code indexing, as it does not perform any of the other listed tasks. 

The~\softname{Research Software Directory} (RSD)~\citesoftware{Spaaks_Research_Software_Directory_2020} is designed to show the impact of RS on scientific community. Its primary objective is to promote RS reuse and encourage proper RS citation to ensure researchers and RSEs get credit for their work. The RSD, by default, is configured to gather RS data from various platforms such as GitHub, \href{https://about.gitlab.com}{Gitlab}, \href{https://orcid.org/}{ORCID}, \href{https://ror.org/}{Research Organization Registry}, \href{https://zenodo.org/}{Zenodo}, \href{https://datacite.org/}{DataCite}, and \href{https://www.crossref.org/}{Crossref}. 
The \softname{RSD} performs bibliometric analysis, citation file checking, and FAIRness aspect checking. 

Software development analytics platforms and toolsets such as \href{https://chaoss.github.io/grimoirelab/}{GrimoireLab}~\citearticle{duenas2021grimoirelab}, 
\href{https://github.com/Software-Improvement-Group/sigridci}{\softname{Sigrid}} and \href{https://github.com/EOSC-synergy/SQAaaS}{\softname{SQAaaS}}~\citearticle{SQAaaS} take a modular approach similar to FAIRSECO employing several tools that collect metadata to compile a quality report of software for their developers. However, none of these tools explicitly consider impact of RS. 

\softname{SQAaaS} focuses on improving RS quality by offering RSEs ready-to-use continuous integration and continuous development pipelines that align with the principles of Open Science~\citesoftware{Naranjo_Delgado_SQAaaS_Web}. It performs license checking, citation file checking, reports on quality of RS. It is integrated with FAIR data assessment tools which follows the FAIRsFAIR route~\citearticle{fairsfair}.

\softname{Sigrid} performs only dependency checking, and the rest of the tasks such as FAIRness aspect, citation file checking, biblometric analysis, license checking are not applicable because the scope of \softname{Sigrid} is on industrial software and its compliance to the industrial standards.
%related checks are focused industrial software and in particular, on ISO/IEC 25010 international standard. 

\softname{GrimoireLab} is an open-source toolkit focusing on data collection, indexing and storage in the \softname{GrimoireLab} database for potential analysis of software development. It provides very powerful and flexible basis for analysis, but requires an additional efforts and expertise to extract and display relevant information. 

Thus, FAIRSECO offers a unique combination in terms of the considered features for assessing impact and quality of RS. 

%GrimoireLab is an open-source toolkit for automatic and incremental data gathering from data source related with contributing to open-source development, automatic gathered data enrichment, and data consumption and visualization.
%Sigrid is a data-driven software intelligence platform, that will analyze all source code derive holistic insights into risks, costs, opportunities on multiple software quality aspects and translate those into business risks and concrete actions.

%%%%%%%%%%%%%%%%%%%%%%%%%%%%%%%%%%%%%%%%%%%%%%%%%%%%%%%
\section{The FAIRSECO framework} \label{fairseco}
%FAIRSECO was developed to meet the requirements of RSEs by providing them with a tool to assess the impact of their RS. 
\begin{figure*}
\centering
\includegraphics[width=1\textwidth]{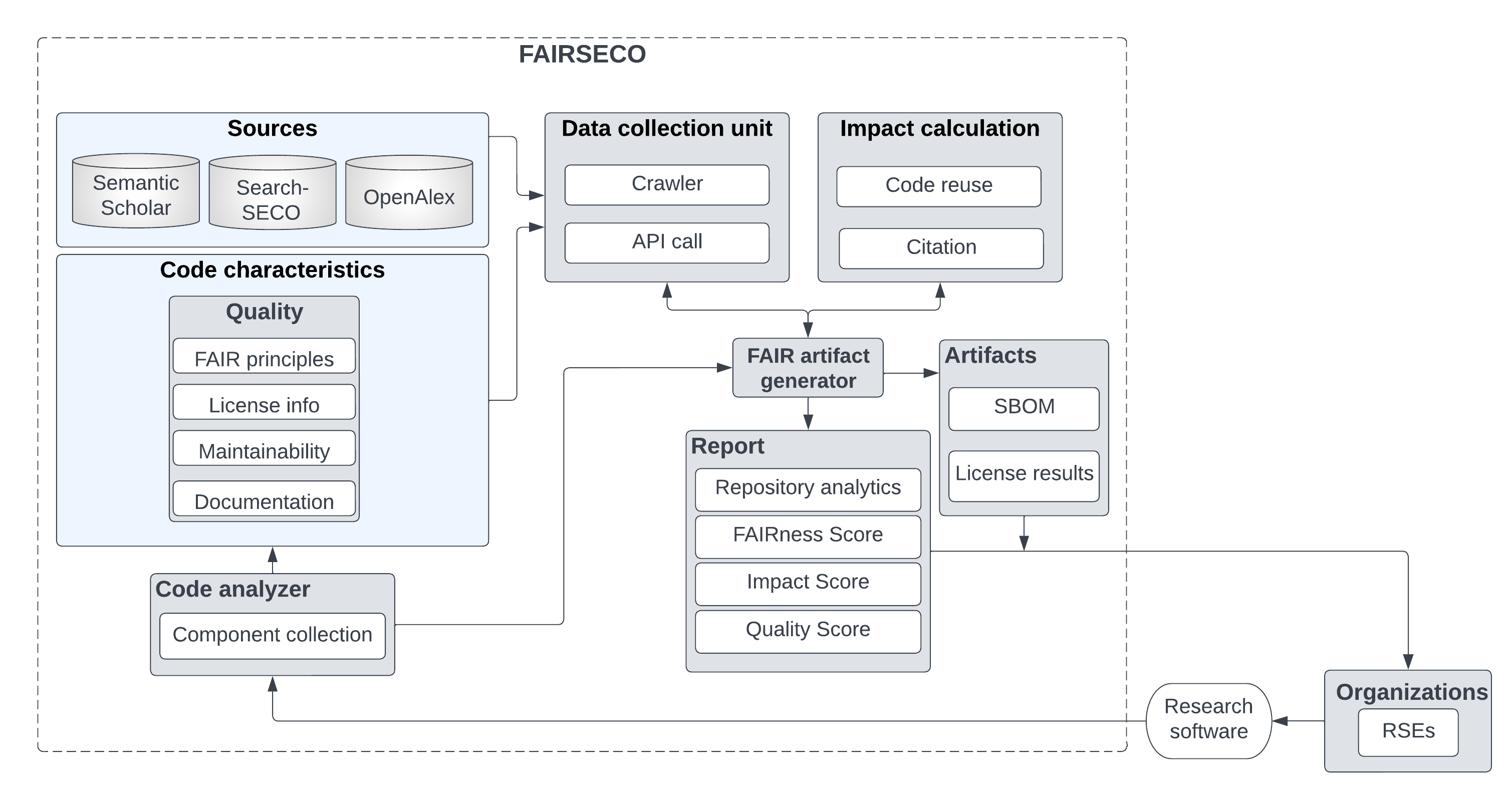}
\caption{FAIRSECO architecture consists of five components: 1) Data collection unit gathers data from different data sources, 2) Impact calculation performs an impact calculation, 3) Code analyzer analyzes the code of the RSE project, 4) Report generates final report related to Quality and Impact, and 5) Artifacts generates two artifacts, such as SBOM, and License results for the RS.}
\label{fig:FAIRSECOArchitecture}
\end{figure*}

\subsection{The FAIRSECO Extensible Software Architecture}

Figure~\ref{fig:FAIRSECOArchitecture} depicts the architecture of the FAIRSECO framework and its components.
The framework consists of five primary components: a Code analyzer, Data collection unit, Impact calculation module, Report, and Artifacts module. All these components are interconnected with a FAIR artifact generator. The FAIR artifact generator will receive input from the Code analyzer, Data collection unit, and Impact calculation module and generates Reports and Artifacts as output. 

The FAIRSECO integrates different sources of information and tools to support RSEs in \emph{becoming more FAIR}, \emph{improving and measuring RS quality}, and \emph{measuring the impact of their RS}. The framework uses the RS's repository such as GitHub as input for checking FAIRness and quality and uses other data sources for measuring impact and quality.

\subsection{Collecting data}
The data collection unit consists of a crawler and an API call component to collect citation data and code reuse details from various sources such as Semantic scholar~\citesoftware{robert}, OpenAlex~\citearticle{priem2022openalex}, and \softname{SearchSECO}.

Besides generating a report, the FAIRSECO framework also creates artifacts, such as:
\begin{itemize}
\item \textbf{SBOM} SBOM is a structured list of third-party software components and libraries included directly or indirectly in the code~\citesoftware{aasim}. It contains information about open-source licenses and version numbers. FAIRSECO will produce an SBOM for the RS using the tool named \softname{SBOM}.
\item \textbf{\softname{Tortellini} Results} 
A detailed report is generated specifically for identifying license violations within the RS. This report provides comprehensive information about the licenses of packages used in the RS, including the primary packages and their dependencies. It includes details such as each package's specific license types and version information. By examining this report, RSEs identify any instances where the RS may violate license agreements, ensuring compliance with licensing requirements and addressing any potential legal or ethical concerns related to the software's usage and distribution.
\end{itemize}

\subsection{Becoming more FAIR}\label{sec:fair}

The FAIRSECO framework assesses the FAIRness score ($S_{fair}$) of RS based on five recommendations for FAIR software~\citearticle{martinez_ortiz_carlos_2020_4310217}, and it is calculated using \softname{howfairis} tool. Explanations for these recommendations are given below:

%\begin{enumerate}\item The \textbf{Findable} criterion checks whether the RS is registered in a community registry such as Zenodo, CodeOcean, FigShare, Software Heritage Archive, etc. Registering RS software makes it easier for others to find it, particularly through the use of search engines such as Google.\item \textbf{Accessible} feature verifies whether the RS uses a publicly accessible repository with version control.\item A quality checklist is utilized by the \textbf{Interoperable} feature to evaluate the interoperability of the RS.\item The RS is considered \textbf{reusable} only if it contains both citation~\citesoftware{Druskat_Citation_File_Format_2021} and license files. This criterion assigns a lower score to the repository without either of these files.
%\end{enumerate}

\begin{enumerate}[label=R\arabic*)]
\item \textbf{Use a publicly accessible repository with version control}: The initial recommendation focuses on verifying that the RS utilizes a publicly accessible repository with version control. This ensures widespread accessibility and encourages transparency. By employing a version control system, RSEs can easily monitor and manage modifications made to the RS, facilitating effective collaboration and traceability.

\item \textbf{Add a license} It is important to emphasize that an appropriate license must accompany the RS. Without a license, even if the RS is publicly available on platforms such as GitHub, the absence of legal permissions makes it impossible for anyone to use it.

\item \textbf{Register your code in a community registry} This recommendation suggests registering RS in a community registry such as Zenodo, CodeOcean, FigShare, Software Heritage Archive, etc. Registering RS makes it easier for others to find it, particularly through the use of search engines such as Google.

\item{\textbf{Enable citation of the software}}
Citation is an integral part of scientific accountability and reproducibility. Citation helps RSEs be recognized for their work. Citation File Format (CFF)~\citesoftware{Druskat_Citation_File_Format_2021} is specifically designed to enable the citation of software.

\item{\textbf{Use a software quality checklist}}
This recommendation proposes examining quality checklists that have been included on \href{https://fair-software.eu/}{fair-software.eu}\footnote{fair-software.eu is a website that promotes FAIR practices in RS development by offering five key recommendations for FAIR software~\citearticle{carlos_2021}. It aims at raising awareness about the importance of making RS FAIR. It provides guidelines and resources for researchers and RSEs to adopt FAIR software principles and best practices.}
\end{enumerate}

The $S_{fair}$ is calculated by combining these recommendations, the RS gets a maximum score of 5, and each recommendation contributes a maximum score of 1 (cf.~\citesoftware{Spaaks_howfairis_2022},~\citesoftware{Verhoeven_fairtally_2021}). We will provide a more detailed explanation in the below subsections on how these scores are calculated.

\subsection{Measuring RS Quality}

%We have taken liberty in interpreting software quality to mean anything that can support RSEs in making their software

The code analyzer in the FAIRSECO framework architecture evaluates the quality of RS. It employs various techniques to analyze the RS, encompassing the evaluation of FAIRness~\citearticle{lamprecht2020towards}, identification of license violations, examination of software maintainability, and checking the presence of documentation. Using $S_{quality}$, RSEs evaluate their project's development and identify improvement areas. The following sections describe the main features contributing to $S_{quality}$.

%%%%%%%%%%%%%%%%%%%%%%%%%%%%%%%%%%%%%%%%%%%%%%%%%%%%%%
\subsubsection{License violation or information}
%Furthermore, licenses play a critical role in developing and distributing open-source software. A license is a legal document specifying the terms and conditions for the software's use, distribution, and modification. By using open-source software, developers agree to the license terms associated with that software.
%The violation of licenses is a widespread problem in society, and it occurs when someone utilizes open-source software without adhering to the terms and conditions of the license, which results in a breach of the license agreement.

License violations have significant consequences for violators and open-source communities~\citearticle{laurent2004understanding}. These violations damage the reputation of a project and discourage further contributions from the community, which have negative long-term effects~\citearticle{Gangadharan}. 

We argue that researchers and RSEs should carefully review the license associated with any open-source software before using it and ensure that they agree to the terms and conditions of that license. This practice can foster the perpetual expansion and prosperity of the open-source ecosystem.

%To mitigate license incompatibility issues, RSEs should adopt a proactive approach. RSEs should carefully review the license associated with any open-source software before using it and ensure that they agree to the terms and conditions of that license. By adhering to proper license compliance, RSEs contribute to the perpetual expansion and prosperity of the open-source ecosystem. This practice fosters trust and collaboration and encourages continued growth and success of open-source communities.

We utilized \softname{Tortellini} for predicting license violations; the FAIRSECO framework helps users become more aware of potential license violations and the potential problems that arise. 

\subsubsection{Maintainability}

Software maintainability refers to the ease with which the operations, such as the addition of new features, obsolete code deletion, and error correction, are carried out~\citearticle{1702216}. It is an aspect of software development that consumes a significant portion of the overall project cost. However, measuring maintainability during the early stages of software development aids in better planning and optimal resource utilization~\citearticle{malhotra2016software}.

The FAIRSECO framework analyzes open and closed issues on GitHub to assess maintainability. It calculates maintainability by calculating the percentage of issues that have been closed since the date on which the software was added to GitHub. RSEs utilize the maintainability score as a guideline to improve their development process and address current open issues more efficiently.

\subsubsection{Documentation}
To be usable, open-source software needs to be sufficiently documented. This helps users understand the functionality and usability of the software. GitHub and Zenodo are platforms that typically include a \readme{} file and documentation  for each software, giving users information about the software. 

The FAIRSECO framework checks whether the RS is adequately documented to ensure users access the necessary documentation. Therefore, FAIRSECO assigns a maximum score of 100 if the RS includes both \readme{} and documentation. 

\subsubsection{Reporting on Quality} 
The code analyzer generates reports that summarize the analysis results. These reports provide RSEs with actionable feedback on the quality of their code. The reports highlight areas that require improvemnt, encompassing aspects such as FAIRness, license information or violations, maintainability, and documentation.  By reviewing these reports, RSEs make informed decisions to enhance the overall quality of their RS.

Finally, we have formulated an equation to determine the $S_{quality}$, incorporating factors such as FAIRness, license violations, maintainability, and documentation: %$S_{quality}$ is calculated based on the following equation: 

\begin{equation}\label{eq:qualityscore}
S_{quality} = \frac{S_{fair} \cdot W_f + S_l \cdot W_l + S_m \cdot W_m + S_d \cdot W_d}{W_{total}}
\end{equation}

%We assume that all weights, 
The all weights, namely, the FAIRness weight $W_f$, license weight $W_l$, maintainability weight $W_m$, and documentation weight $W_d$ are four distinct constant assigned to each factor. The total weight $W_{total}$ is calculated by summing up these weights $W_{total} = W_f + W_l + W_m + W_d$. Currently, $S_{quality}$ is determined using weighted criteria; however, future work will investigate the justification behind these weights. The scores used in~\eqref{eq:qualityscore} are calculated as follows:
\begin{itemize}
    \item The $S_{fair}$ is calculated by multiplying the \softname{howfairis} tool's output (0-5) by 20.  
    \item The license score $S_l$ is calculated using the license information provided by the \softname{Tortellini} tool as 
\begin{equation}\label{eq:licensescore}
S_l = {F_l}^{\log_2(1 + N_l)} \cdot 100
\end{equation}

where $F_l$ is the fraction of licenses that are not violated, and 
$N_l$ is the total number of licenses.
    \item The maintainability score $S_m$ is calculated as follows:
\begin{equation}\label{eq:maintainscore}
S_m = \frac{100 \cdot C_i} {N_i}                                                                                                
\end{equation}
where $C_i$ is the number of closed issues, and $N_i$ is the total number of issues.
\item The documentation score $S_d$ is determined based on the presence of \readme{}, and documentation files. The score is assigned a value of 50 if there is a documentation directory and another 50 if there is a \readme{} file. These scores are then combined to obtain the overall documentation score.

\end{itemize}
Thus, a score between 0 and 100 will be assigned to the RS based on Equation~\eqref{eq:qualityscore}.
The calculations described above represent our current FAIRSECO approach, considered a starting point. However, we plan to enhance the formulas by incorporating more detailed and higher-quality metrics. The framework is designed to be sufficiently generic to accommodate these expansions.

%%%%%%%%%%%%%%%%%%%%%%%%%%%%%%%%%%%%%%%%%%%%%%%%%%
\subsection{Measuring RS Impact}

One of the largest challenges in providing insight into RS impact is where and how to report this data. There are indicators such as GitHub stars, inclusions of a package in dependency trees, citations to articles, mentions in news items, the number of issues posted by non-community members, etc. Gathering and interpreting these data is challenging. For FAIRSECO, currently, we have implemented two impact measures: citations and code reuse. Furthermore, researchers would likely hesitate to use poor-quality RS that introduces bugs, performance and maintenance issues, etc., because of potentially erroneous results produced by the RS and higher costs for fixing its technical debt~\citearticle{debt:2016}. Thus, we consider the quality of RS as an additional factor that influences $S_{impact}$, along with citations and code reuse. This intuition is in line with recent efforts of policymakers regulating long-term plans for maintaining RS~\citearticle{martinez:2023}.

\subsubsection{Citation} In academic writing, the practice of citing relevant literature to establish connections with other works in the field is considered essential and common practice~\citearticle{arsyad2018review}. Recognizing the significance of this practice, FAIRSECO offers valuable assistance to its users by facilitating the rapid and efficient retrieval of citation data for their RS. This is accomplished by gathering information from APIs of the open catalogues of scientific publications (in particular, OpenAlex and Semantic Scholar), enabling users to conveniently obtain the necessary citations for their RS. Moreover, it collects information about the scientific fields in which the RS is cited. It generates a radar citation graph~\citearticle{haghnazar2021visualization} illustrating how the RS is connected across different scientific fields. A screenshot is provided in Figure~\ref{fig:Impact}.

\subsubsection{Code Reuse} For this, the FAIRSECO framework utilizes \softname{SearchSECO}: it feeds a repository link to \softname{SearchSECO}, which then parses all the source code from the RS project and calculates a hash for an abstract version of each method in the work. Subsequently, \softname{SearchSECO} looks in its database for occurrences of the method in the worldwide software ecosystem. If found, there are two possible outcomes:
\begin{enumerate}
    \item (A fraction of) the RS is being reused by other parties, showing its impact. No further action is needed if the other software uses the RS's intellectual property correctly.
    \item The RS project is reusing other software. While this does not lead to impact, it does provide an opportunity to ensure that the code in the RS project is used correctly with correct references and that the reused methods do not contain any vulnerabilities, which is a feature of \softname{SearchSECO}.
\end{enumerate}

\subsubsection{Reporting on Impact}
In addition to citations and code reuse, we include $S_{quality}$ as a factor influencing the Impact score. 
%Poor quality of RS would include bugs, lack of proper testing, performance issues, maintainability challenges; this would make it  prohibitive for researchers to use RS in their work in a fear of unsound results, non working software. Thus, we 
Similarly to the $S_{quality}$, the $S_{impact}$ is determined by applying constant weights to factors such as the number of citations $N_c$, code reuse in other projects ($N_r$) as well the $S_{quality}$ obtained from Equation~\eqref{eq:qualityscore}:
\begin{equation}\label{eq:impactscore}
S_{impact}= \frac{N_c \cdot W_c + N_r \cdot W_r + S_{quality} \cdot W_q}{W'_{total}}
\end{equation}
where citation weight $W_c$, reused weight $W_r$, quality weight $W_q$ are constants.
$W'_{total}$ is calculated by summing up these weights: $W'_{total} = W_c + W_r + W_q$.

Note that these constant weights can be changed after an expert evaluation and the impact measurement metric can be extended with more elaborate metric.

\section{The FAIRSECO framework in Action}\label{toolaction}

This section overviews how RSEs utilize FAIRSECO to generate reports on FAIRness, Quality, and Impact. Generally, RSEs will want to use the FAIRSECO framework every couple of months or when they wish to provide a report on the impact of their repository. The history feature of the FAIRSECO infrastructure supports this, as the RSEs show the improvements they made (or did not make) regarding the impact and citation of their software. For a complete manual to operate the FAIRSECO GitHub Action, please see the documentation at 
\href{https://github.com/SecureSECO/FAIRSECO#readme}{\readme{}}.

RSEs face challenges in monitoring the usage of their software, tracking academic papers that reference it, and identifying software built upon it. With the FAIRSECO framework, RSEs easily access information on license violations, impact, citation details, and the $S_{quality}$ of their RS. This eliminates the need to identify such details, thereby saving time and effort manually. The report generated by the FAIRSECO framework increases the chances of other researchers citing the work of RSEs.

To illustrate the functionality of the FAIRSECO framework in a RS context, we have chosen the \softname{Mcfly} tool~\citesoftware{van_Kuppevelt_mcfly_deep_learning_2022} from GitHub. The following are the pages or tabs within the FAIRSECO framework, each with its respective explanation:

\begin{figure}[ht]
\centering
\includegraphics[width=1.0\linewidth]{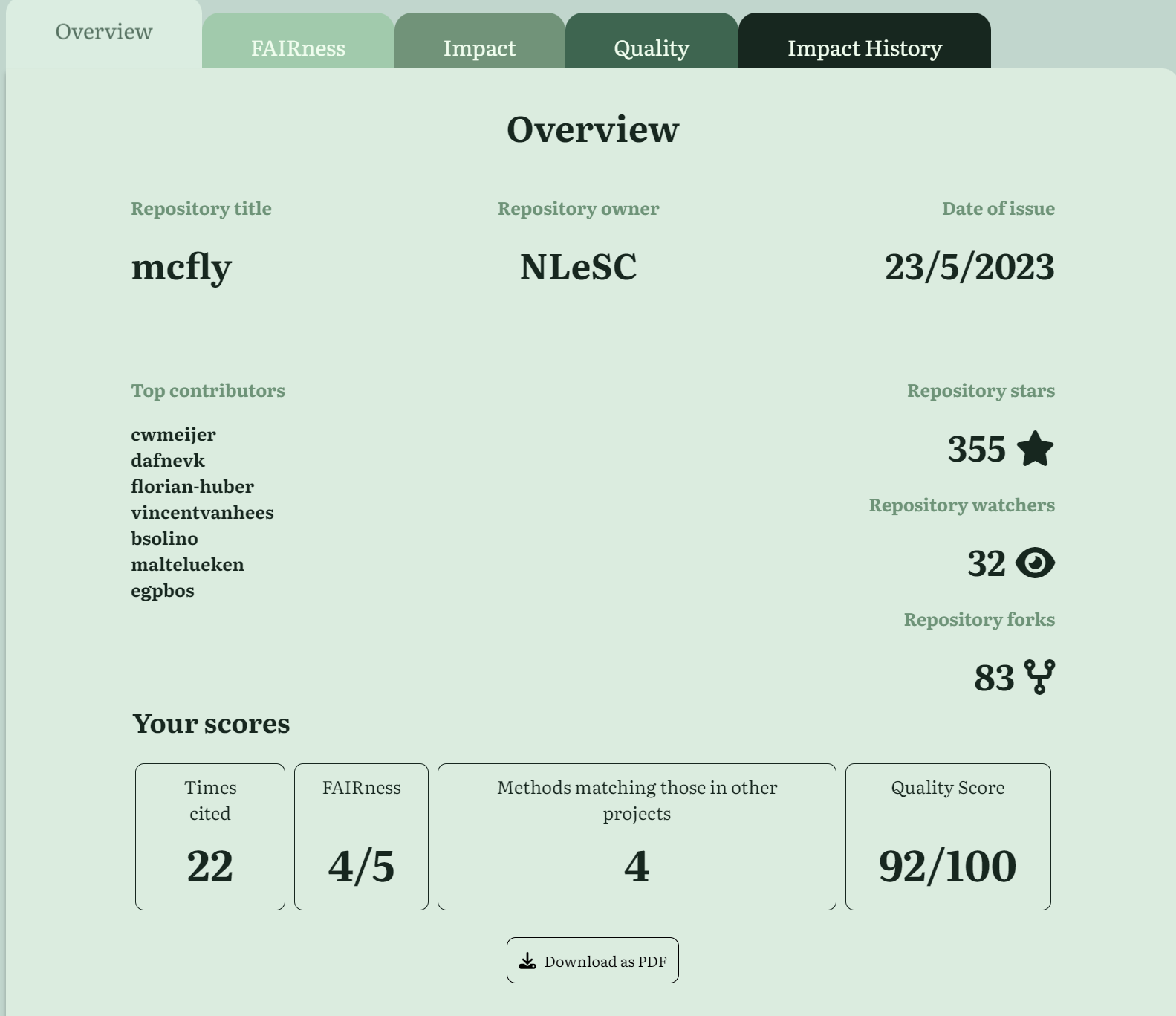}
\caption{ Overview tab: This screenshot displays a summary of the FAIRSECO report, including the number of citations, $S_{fair}$, matching methods, $S_{quality}$, and several metrics from GitHub. For a more detailed understanding of the $S_{quality}$, please refer to Section \ref{toolaction}}
\label{fig:FAIRSECO-Overview}
\end{figure}

\noindent\textbf{1) Overview tab}: In Figure~\ref{fig:FAIRSECO-Overview}, we provided a screenshot of the ``Overview'' tab of the framework, which displays various details about the evaluated tool. These details include the repository title, owner, issue date, number of stars, watchers, and forks. Detailed explanations of each of these terms are given below:
\begin{enumerate}
    \item 
\textbf{Repository title}: Repository title refers to the repository's name, which is evaluated using the FAIRSECO framework.
\item
\textbf{Repository owner}: The repository owner is the name of the individual or organization who owns the repository on GitHub. In case the repository has been forked, it shows the user's name who forked the primary repository.
\item
\textbf{Date of issue}: The issue date refers to when the repository was assessed using the FAIRSECO framework.
\item
\textbf{Repository stars}: The Repository stars refer to the number of stars the repository receives on GitHub.
\item
\textbf{Repository watchers}: The number of users who have added the repository to their watchlist on GitHub is added to the Repository watchers list.
\item
\textbf{Repository forks}: The Repository forks refers to the number of times the repository has been forked on GitHub.
\end{enumerate}

Moreover, the FAIRSECO framework gives further information, including citation counts, FAIRness value, method matching with other projects, and $S_{quality}$ for the evaluated repository. Detailed explanations of these terms are provided in the FAIRSECO framework's next tab. Additionally, there is an option to download this report as a PDF.

\noindent\textbf{2) Citation tab}:
An overview of the research papers that have cited RS is presented in the citation tab. The list of papers includes their title, author's names, description indicating the source from which the paper is collected, publication date, DOI, and a link to the paper.

\noindent\textbf{3) FAIRness}:
The report on FAIRness is generated by analyzing the results of the \softname{howfairis} tool, and it is presented in the FAIRness tab. 
For this example tool, it has received a $S_{fair}$ of 4 out of 5, because it did not meet all the quality criteria outlined in the 5 recommendations for FAIR software.

\noindent\textbf{4) License violation}:
The FAIRSECO employs the \softname{Tortellini} tool to provide a comprehensive account of license violations, informing the original author of the repository or the user about any infractions. The license violations tab displays the information regarding license violations.

\begin{figure}[ht]
\centering
\includegraphics[width=1.0\linewidth, ]{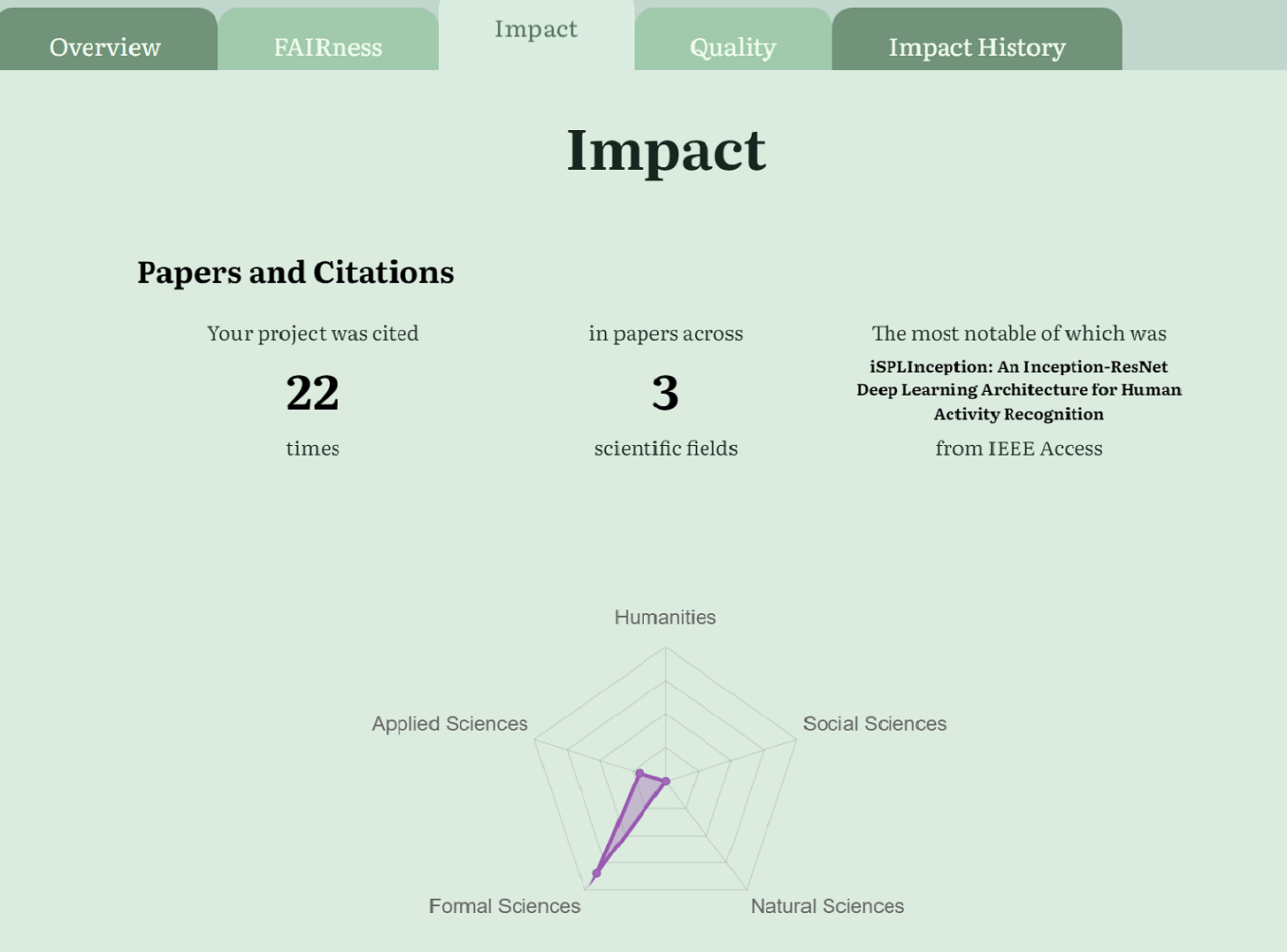}
\caption{Impact tab: The Impact tab screenshot presents an overview of the RS \softname{Mcfly} impact. It has been cited a total of 4 times, with two of those citations coming from different fields. Additionally, the screenshot includes a radar graph that illustrates the software's citation frequency across various fields. For a more detailed understanding of this tab, please refer to Section~\ref{toolaction}.}
\label{fig:Impact}
\end{figure}

\noindent\textbf{5) Impact}: As shown in Figure~\ref{fig:Impact}, the Impact section of FAIRSECO displays various statistics about using the software in scientific research. It includes information about the frequency of citations the software has received across different fields, visually represented in a radar chart showcasing these fields' distribution. Additionally, it provides information on the most significant scientific paper that cited the software based on the number of citations and the journal's reputation.

\begin{figure}[ht]
\centering
\includegraphics[width=1.0\linewidth]{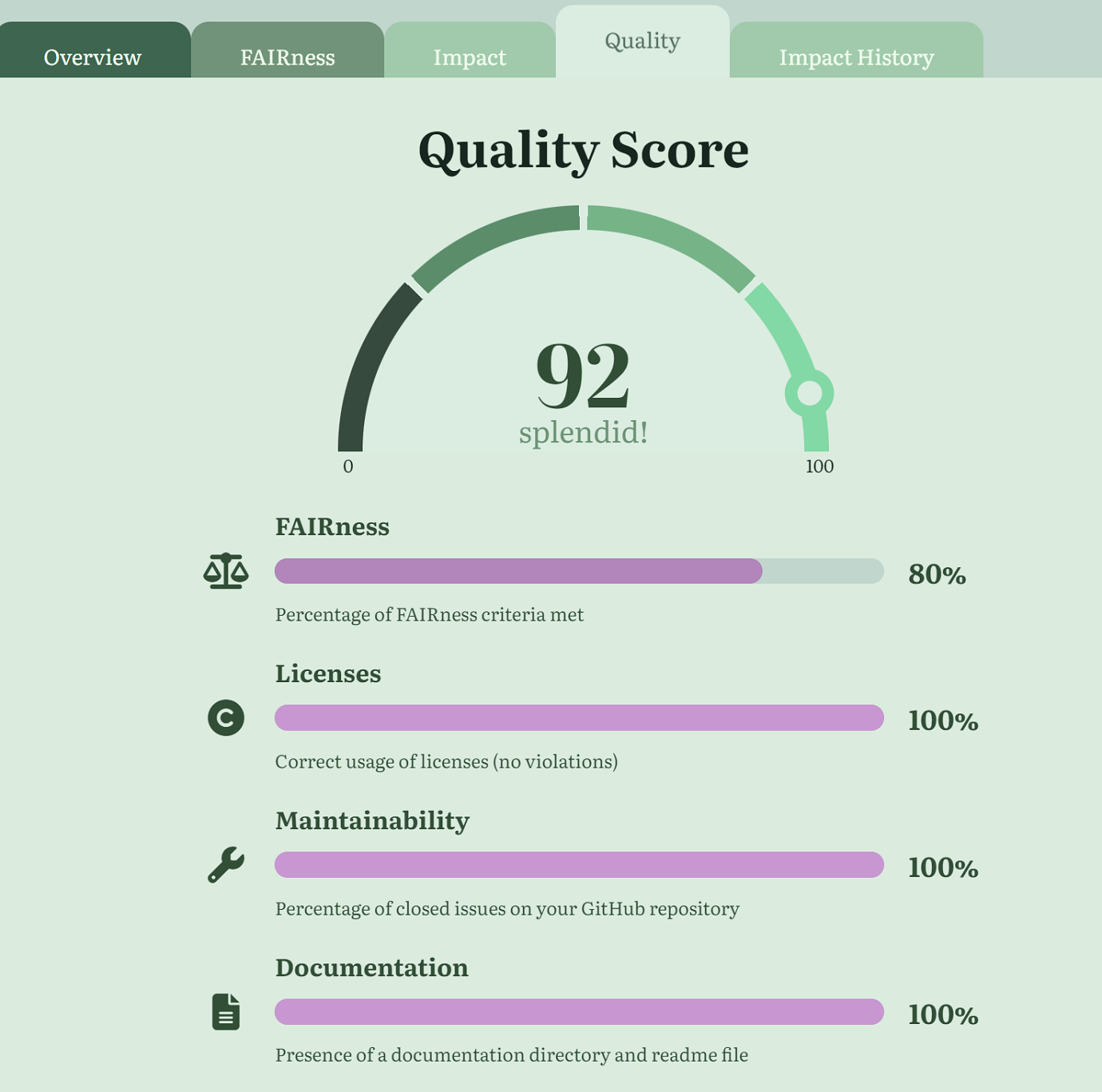}
\caption{ Quality Score tab: This screenshot shows an overview of the generated FAIRSECO $S_{quality}$. For a complete explanation of the $S_{quality}$, see Section~\ref{toolaction}}.
\label{fig:QualityScore}
\end{figure}

\noindent\textbf{6) Quality Score ($S_{quality}$}): The $S_{quality}$ tab is illustrated in Figure~\ref{fig:QualityScore}. The FAIRSECO assigns a $S_{quality}$ to each project and explains how this score was calculated. Four key metrics are considered, including the percentage of adherence to FAIRness principles, the extent to which the software complies with the license requirements, the percentage of closed issues on GitHub, and the presence of \readme{} and documentation files. This score doesn't reflect the actual quality of a project's code but rather indicates how well the project follows the best practices of software engineering. The example repository received a $S_{quality}$ of 92\% out of 100 because it did not pass the quality checklist. Thus, it receives a FAIRness value of 4 out of 5. 

%%%%%%%%%%%%%%%%
\noindent\textbf{7) Impact History}: The impact history tab provides a historical overview of the impact of RS by displaying information from previous runs of the FAIRSECO, if available. This feature allows RSEs to track how the impact of their software has evolved. The overview section shows the changes in $S_{quality}$, number of citations, reused methods, and $S_{fair}$ since the last run. Ideally, RSEs will observe an increase in these metrics over time. The graph below visually represents this data, including runs before the most recent one. 

\begin{table*}[ht]
\scriptsize
\centering
\caption{Evaluation table: This table compares selected tools using the FAIRSECO framework. See Section~\ref{sec:fair} for the definition of R1-R5.}
\label{evaluation table}
\begin{tabular}{|l|l|l|l|l|l|r|r|r|r|r|}  
\hline
\multicolumn{1}{|c|}{Tools} & \multicolumn{10}{c|}{Feature}                                \\\hline
%\cline{2-11}&\multicolumn{3}{|c}{} & \multicolumn{2}{|c|}{Reusable}& \multicolumn{5}{c|}{} \\\hline

& \turntitle{R1} & \turntitle{R2} &  \turntitle{R3} & \turntitle{R4} & \turntitle{R5} & \turntitle{FAIRness}  & \turntitle{Licenses} & \turntitle{Maintainability}  & \turntitle{Documentation} & \turntitle{Quality Score}   \\\hline

% & \multirow{Findable} & \multirow{Accessible} & \multirow{Interoperable} & \multicolumn{2}{l|}{Reusable}    & \multirow{FAIRness} & \multirow{Licenses} & \multirow{Maintainability} & \multirow{Documentation} & \multirow{Quality score} \\ \cline{5-6}    &                           &                             &                                & \multicolumn{1}{l|}{\begin{tabular}[c]{@{}l@{}}Citation\\ file\end{tabular}} & \begin{tabular}[c]{@{}l@{}}License\\ file\end{tabular} &                           &                           &                                  &                                &                                \\ \hline
\softname{Kernel Tuner}   & \passed & \passed  &  \passed & \passed & \passed  & 100\% & 100\%  & 100\% &50\%  & 94\%  \\ 
\hline
\softname{Spec2Vec} & \passed & \passed  & \passed & \passed & \passed & 100\% & 100\%  & 100\% &50\%  & 94\%   \\ 
\hline
\softname{ESMValTool}  & \passed   & \passed   & \passed  & \passed & \failed   & 80\%  & 100\% &100\%  &50\%  & 86\%\\ 
\hline
\softname{GPT index}  & \failed    & \passed     & \passed   & \passed & \failed & 60\%  & 100\%   & 100\% & 100\% & 85\%    \\ 
\hline
\softname{CFF-Converter-Python}   & \passed & \passed    & \passed & \passed & \passed    & 100\% & 19\%  & 100\% & 100\%   & 78\% \\ 
\hline
\end{tabular}

\end{table*}

\begin{table*}[ht]
\scriptsize
\centering
\caption{Impact Score Table: Impact Score table shows the influences of various factors on the overall Impact Score}
\label{tab:impactTable}
\begin{tabular}{|l|r|r|r|r|} 
\hline
\multicolumn{1}{|c|}{Tools} & \multicolumn{4}{c|}{Feature}                                   \\ 
\cline{2-5}
\multicolumn{1}{|c|}{} & \# citations & Reused in other Projects  & Quality Score & Impact Score  \\ 
\hline
\softname{Kernel Tuner} & 17 &  5 &   94\%
& 60\% \\ 
\hline
\softname{Spec2Vec} & 4 & 5 & 94\% & 56\% \\ 
\hline
\softname{ESMValTool} & 40 & 4& 86\%& 61\%\\ 
\hline
\softname{GPT index} & 0 & 3 & 85\% & 50\%\\ 
\hline
\softname{CFF-Converter-Python} & 0& 5 & 78\% & 46\% \\ 

\hline
\end{tabular}
\end{table*}
%%%%%%%%%%%%%%%%%%%%%%%%%%%%%%%$
\section{Evaluation} \label {evaluation}
%EVALUATING TOP 5 PROJECTS
Table~\ref{tab:comparisonTable} compares five tools based on their performance concerning features such as Recommendation 1 (R1), Recommendation 2 (R2), Recommendation 3 (R3), Recommendation 4 (R4), Recommendation 5 (R5), license violation, maintainability, and documentation. The FAIRSECO framework calculates a $S_{quality}$ for each evaluated tool based on these features, also provided in the table. Each of these tools such as \softname{Kernel Tuner}~\citesoftware{van_Werkhoven_Kernel_Tuner_2019}, \softname{Spec2Vec}~\citesoftware{Huber_spec2vec}, \softname{ESMValTool}~\citesoftware{Andela_ESMValTool_2023}, \softname{GPT index}~\citesoftware{Liu_LlamaIndex_2022}, 
and \softname{CFF-Converter-Python}~\citesoftware{Spaaks_cffconvert_2021} were evaluated using the FAIRSECO tool. Among the five tools compared,  \softname{Kernel Tuner} and \softname{Spec2Vec} met the FAIRness criteria and had a higher $S_{quality}$ than the other tools. \softname{Kernel Tuner} stands out in its FAIRness as it is registered in Zenodo with a DOI, making it easily findable. Additionally, it is publicly accessible through GitHub and downloaded using standard protocols such as HTTPS. Moreover, \softname{Kernel Tuner} has citation and license files, further enhancing its FAIRness. Therefore, \softname{Kernel Tuner} scores a perfect 5 out of 5 regarding FAIRness criteria. According to our analysis, it does not violate any licenses. \softname{Kernel Tuner} demonstrates effective maintenance by actively closing issues on GitHub, which is properly documented through documentation files and a \readme{} file. The \softname{ESMValTool} does not meet the quality checklist mentioned in 5 recommendations for FAIR software and receives a lower $S_{quality}$ than \softname{Kernel Tuner} and \softname{Spec2Vec}. \softname{GPT index} scored lower in the R1 and R5 categories but performed well in R3, R4 and R5. \softname{CFF-Converter-Python} scored relatively low in the $S_{quality}$. Still, it performed well in all five features, scoring less due to issues with the version of the PyPI package, resulting in a license violation.
 
In summary, \softname{Kernel Tuner} and \softname{Spec2Vec} stand out as highly reliable and FAIR tools, since while \softname{ESMValTool}, \softname{GPT index}, and \softname{CFF-Converter-Python} show strengths in specific areas but may require some attention to enhance their overall performance.

Table~\ref{tab:impactTable} gives the $S_{impact}$ for selected tools. As per Equation~\eqref{eq:impactscore}, $S_{impact}$ is calculated based on three factors: the number of citations, the number of methods reused, and the $S_{quality}$. For example, the tool \softname{Kernel Tuner} has 17 citations, has been reused in 5 other projects, and has a $S_{quality}$ of 94\%. Using Equation~\eqref{eq:impactscore}, the $S_{impact}$ for \softname{Kernel Tuner} is 60\%.

The FAIRSECO framework provides a comprehensive report on the quality and impact of RS on research. RSEs utilize this report to identify areas where improvements are needed and make necessary changes to their software.

%%%%%%%%%%%%%%%%%%%%%%%%%%%%%%%%%

\section{Conclusion and Future Work}
\label{conclusion}
% lessons learned 
 % future work
RS is becoming increasingly valued in the research ecosystem, leading to the evolution of international and national policy practices to reflect its importance. The FAIRSECO framework, with its collection of tools, is designed to report on the quality and impact of RS. FAIRSECO framework automatically measures the FAIRness of the RS 
using five recommendations for FAIR software. This is achieved by conducting several checks to ensure that the RS is registered in a community registry, making it available in a public repository, using a quality checklist, encouraging the use of the CFF to facilitate appropriate citation, and including a license file to enable reuse by other researchers.

To assess the quality of the software, the FAIRSECO framework calculates a Quality Score based on several factors, including FAIRness, license violation, maintainability, and documentation. It also generates an SBOM for the given repository. In addition, FAIRSECO provides insights into the software's impact on the RS ecosystem by gathering citation data and tracking its reuse by other software. The Impact Portal generated by FAIRSECO enables RSEs to assess the success of their software and gain insights into its impact. By showing this, we hope to raise awareness that software is valuable research output. 

From the FAIRSECO report, RSEs understand the areas they must focus on to make their RS available worldwide and improve its quality. This allows users to quickly find the RS they need for their purposes, which increases its popularity. When an RS gains more users, its impact on research helps RSEs receive the credit they deserve. 

In the future, we plan to:
\begin{enumerate}
\item 
\textit{Evaluate with RSEs} -- As part of the design study, we plan to present the framework to RSEs and gather their feedback for evaluation, and further development.
\item
\textit{Promote the FAIRSECO GitHub Action} -- We want to encourage RSEs to use the FAIRSECO framework. We will organize events for them and present the framework at various FAIR software events.
\item
\textit{Create a Live Dashboard for FAIRSECO Impact} -- The current GitHub action is a manually triggered task. We imagine that RS projects need a continuous impact dashboard, but we will evaluate this first with the intended users.
\item 
\textit{Expand the functionality of the FAIRSECO tool to support other platforms} -- Currently, the framework is designed to process projects hosted on GitHub, primarily through a GitHub action. However, we aim to explore the potential of extending FAIRSECO's capabilities to include repositories or tools hosted on various other platforms. This expansion will allow for a broader range of applications and make FAIRSECO more versatile and accessible to users across different development environments. 
\item 
\textit{Expand documentation score} -- At present, FAIRSECO's functionality is limited to checking the presence of documentation in the repository. However, in the future, our aim is to use Natural Language Processing techniques to verify complexity and extensiveness of the documentation.
\item 
\textit{Introduce a dependency score} --
Currently, our code reuse score is restricted to the results obtained from \softname{SearchSECO}. To enhance the effectiveness of the code reuse analysis, we intend to also incorporate a dependency score by detecting which other libraries and tools reuse a particular RS. This additional score will strengthen the outcomes of the code reuse analysis significantly.
%In the future, we plan to calculate a dependency score by detecting which other libraries and tools reuse a particular RS. Using this score combined with the code reuse score will provide a comprehensive insight into the reuse of the RS.
\item 
\textit
{Score Selection and Weighting} -- There are different stakeholders in the development of a research software project, such as the scientists who need it, the funding agency, and the RSEs. Each of these is interested in different scores; the scientists want to know who uses the software in their network, the funding agency wants to know whether the software is developed so it does not suffer too much from so-called software rot, and. the RSEs wish to know how many open bugs there are and their severity. For this reason different scores are required to provide insight into the different qualities of a research software project. Furthermore, it could be that composite scores are needed to rapidly provide insight into the software's qualities.
%\textit{Selection and experiment with weights of the scores} -- We will consider different distributions of weights, for example, based on commit activities to focus on main developers, number of developers, etc. To that end, our prime candidate for gathering such data is \softname{GrimoireLab}. We plan to assess which weights are relevant for which stakeholders. Using that information will further enable us to compose separate FAIRness score, should there be a need for those.
\item 
\textit{Expand the notion of impact} -- Ideally, a term \emph{impact} would mean how the RS has influenced and advanced research that it was used in~\citearticle{impact:2013}. However, this is difficult to measure with the data available for RS. Most of the research uses quantitative impact metrics from bibliometric analysis, namely, the number of citations and code reuse (which applies to the software only). We have included these and a Quality Score as yet another factor influencing impact. In the future, we want to consider a broad notion of \emph{impact} that includes both academic and socio-economical aspects, such as the number of grants and information regarding people hired to maintain the RS.

%The latter data is challenging to collect because of the GDPR and other privacy concerns.
\end{enumerate}

By extending FAIRSECO with these features, we strive to guide researchers and RSEs to adapt high quality standards for their RS, and provide more accurate insight into (potential) impact of their RS.

%measuring impact of RSRSEs will receive continuous support in improving the FAIRness of their RS projects. This lead to better recognition and acknowledgment of their work within the research community.

\section*{Acknowldegment}
We thank the eScience Center for funding this Project (NLESC.CIT2021.002). We also thank the students who supported the software development of the FAIRSECO framework: A. Aydin, B. Hageman, B. Lankhorst, J. Hendriksen, Q. Donkers, R. Schouten, T. Bolhuis, \& V. Bykova.

\balance

\renewcommand\refname{Article References}

\renewcommand\refname{Software References}

%\bibliographystylearticle{myIEEEtran}
%\bibliographyarticle{refs}
%\bibliographystylesoftware{myIEEEtran}
%\bibliographysoftware{refs}

\end{document}